\documentclass[conference]{IEEEtran}
\usepackage{tikz,graphicx,xcolor}
\usepackage{gensymb}
\usepackage{textcomp}
\usepackage{float,dblfloatfix}
\usepackage{verbatim,array,enumitem}
\usepackage{algorithm,algpseudocode,breqn,amsmath}
\usepackage{adjustbox,diagbox,threeparttable,tablefootnote,colortbl}
\usepackage{booktabs,tabularx,tabu,multirow,multicol,hhline}
\usepackage{caption,subcaption}
\PassOptionsToPackage{hyphens}{url} 
\usepackage{hyperref}
\usepackage[capitalise,noabbrev]{cleveref} 
\crefname{appsec}{Appendix}{Appendices} 
\usepackage[misc]{ifsym}
\usepackage[utf8]{inputenc}
\usepackage{booktabs}
\usepackage{lipsum}
\usepackage{tabularray}
\usepackage{titlesec}
\usepackage{longtable}
\usepackage{mdframed}
\usepackage{tcolorbox}
\usepackage{multicol}

\usepackage{listings}
\usepackage{xcolor}

\lstdefinestyle{python}{
    language=Python,
    basicstyle=\ttfamily\footnotesize,
    keywordstyle=\color{blue},
    stringstyle=\color{green!40!black},
    commentstyle=\color{gray},
    showstringspaces=false,
    breaklines=true,
    frame=single,
    captionpos=b,
    numbers=left,
    numberstyle=\tiny\color{gray},
    stepnumber=1,
    numbersep=5pt,
}

\titleformat{\subsubsection}
  {\normalfont\normalsize\itshape}{\thesubsubsection}{1em}{}

\titlespacing*{\subsubsection}{0pt}{1.25ex plus 1ex minus .2ex}{0.75ex plus .2ex}

\newcommand{\name}{\textit{EtherBee}}
\title{\name: A Global Dataset of Ethereum Node Performance Measurements Coupled with Honeypot Interactions and Full Network Sessions}

\author{
\IEEEauthorblockN{Scott Seidenberger and Anindya Maiti}
\IEEEauthorblockA{University of Oklahoma \\
seidenberger@ou.edu, am@ou.edu}
}

\begin{document}

\maketitle

\begin{abstract}

We introduce \textit{EtherBee}, a global dataset integrating detailed Ethereum node metrics, network traffic metadata, and honeypot interaction logs collected from ten geographically diverse vantage points over three months. By correlating node data with granular network sessions and security events, \textit{EtherBee} provides unique insights into benign and malicious activity, node stability, and network-level threats in the Ethereum peer-to-peer network. A case study shows how client-based optimizations can unintentionally concentrate the network geographically, impacting resilience and censorship resistance. We publicly release \textit{EtherBee} to promote further investigations into performance, reliability, and security in decentralized networks.

\end{abstract}

\section{Introduction and Related Work}

Web3 networks are becoming critical public infrastructure, and Ethereum is among the most widely used and robust public smart contract platforms. As the network grows in size and complexity, understanding the performance, behavior, and risk characteristics of Ethereum nodes is a key component to ensuring the blockchain’s security, reliability, and scalability. Nodes are the core components that validate transactions, execute smart contracts, and maintain global consensus; however, the decentralized nature of blockchain networks complicates holistic data collection and analysis of these nodes. Existing datasets often fail to capture the full breadth of node activity when striving for global coverage, granular metrics, and long observation periods \cite{surveyofdatasets, wang2024exgraph, chen2019dataether}.

Recent work has examined Ethereum node performance and decentralization, showing that the effectiveness of the network can shift over time due to major events like client software bugs, and that a small number of large entities often control most validators \cite{grandjean2023ethereumproofofstakeconsensuslayer, cortes2021discovering}. Furthermore, economic incentives may unintentionally promote centralization \cite{cortes2023hidden}. Other studies focus on measuring the Ethereum networking stack through active monitoring \cite{ethereumNetworkPeers, wang2021ethna, gencer2018decentralization}, yet most publicly available datasets prioritize on-chain activity rather than node or security-focused data. Meanwhile, multimodal cyber threat intelligence has proven essential to uncovering sophisticated attacks, especially when analyzing logs, network telemetry, and threat feeds in tandem \cite{almohannadi2018cyber, mahboubi2024evolving}.

To address these gaps, we present \name, a first-of-its-kind global dataset that integrates detailed Ethereum node metrics, network traffic metadata, and honeypot interaction logs. By capturing both benign and malicious activity across diverse geographic vantage points, \name~facilitates a deeper understanding of peer-to-peer (P2P) communication, node stability, and network-level threats. To the best of our knowledge, no existing resource provides this level of multimodal detail that covers both operational node data and honeypot collections. In summary, our contributions are: (1) \name, a dataset with fine-grained metrics from 10 vantage points over three months; (2) a scalable methodology for collecting, storing, and transforming the data; and (3) analyses that generate new research questions around performance, reliability, and security for decentralized networks.

\section{Data Collection}

We deployed ten AWS servers across five global regions (North America, Europe, Middle East, Asia, and South America). In each region, one server ran Lighthouse \cite{lighthouse} (Consensus Layer), Nethermind \cite{nethermind} (Execution Layer), plus honeypot and packet capture software; a second server ran only honeypot and packet capture. All three data streams—network sessions, node metrics, and honeypot logs—were indexed into a six-node Elasticsearch cluster (227\,TB storage, 132\,GB heap), yielding 30.1\,TB of primary data across 1,419 shards and 33.7\,billion documents. We used a WireGuard-based mesh to encrypt inter-server traffic, and afterwards exported raw Elasticsearch data to tabular formats for public release. This final, time-synchronized dataset enables correlating network-level events, node-level performance, and cyberattacks at global scale.

\subsection{Network Traffic Capture}
Using Arkime \cite{arkime}, we indexed start/end times, IP addresses (with geolocation), ports, and byte counts for every connection. Over three months (Aug--Nov 2024), 94{,}659 unique IPs accessed our P2P ports, yielding 20.97\,TB of metadata. Since Ethereum mainnet typically has 6--10k persistent beacon nodes \cite{monitoreth}, our dataset encompasses both stable peers and transient endpoints.

\subsection{Node Metrics and Honeypot Logs}
We instrumented Lighthouse (including its attestation simulator\footnote{https://blog.sigmaprime.io/attestation-sim.html}) and Nethermind to capture chain health (slots since best block, attesters, aggregator participation), node performance (latencies, errors, reorganizations), and peer/network stability (disconnects, discovery events). Simultaneously, a diverse honeypot suite \cite{tpot} logged 133.8\,million inbound attacks, from port scans to CVE-specific exploits. Of these, 58\% were DDoS-related, closely matching Cloudflare Radar’s 57\% \cite{radar}. Collecting these security events alongside blockchain metrics offers a unique multimodal lens on both benign and hostile interactions targeting Ethereum nodes.

\section{Case Study}

\subsection{Motivation \& Setup}

\textbf{Peer Pruning and Geographic Centralization.} In latency-sensitive networks like Ethereum, client software scores peers based on metrics such as RTT \cite{krajsa2011rtt,hu2012towards}, actively pruning those with poorer scores. While this can enhance propagation efficiency, existing research suggests it may inadvertently favor geographically closer peers, reducing overall decentralization \cite{mao2020perigee, gencer2018decentralization}. To evaluate this, we analyzed three months of P2P sessions from \name, mapping peer IPs to geographic coordinates (via MaxMind GeoLite2) and computing a data-weighted “focal point.” Each peer’s weight corresponded to its total bytes exchanged, reflecting its relative influence on the node’s connectivity.

\textbf{Weighted Geographic Focal Point Calculation.} For $N$ peers with latitude $\phi_i$, longitude $\lambda_i$, and weight $w_i$, we convert each location to 3D Cartesian coordinates on the unit sphere:
\[
x_i = \cos(\phi_i)\cos(\lambda_i), \quad
y_i = \cos(\phi_i)\sin(\lambda_i), \quad
z_i = \sin(\phi_i).
\]
We then compute:
\[
X = \tfrac{\sum w_i x_i}{\sum w_i}, \quad
Y = \tfrac{\sum w_i y_i}{\sum w_i}, \quad
Z = \tfrac{\sum w_i z_i}{\sum w_i},
\]
and finally:
\[
\phi_c = \arctan2\bigl(Z, \sqrt{X^2 + Y^2}\bigr), \quad
\lambda_c = \arctan2(Y, X).
\]
The resulting $(\phi_c, \lambda_c)$ indicates the daily spatial centroid of peer connectivity.

\subsection{Key Findings}

\begin{figure}[t]
    \centering
    \includegraphics[width=\linewidth]{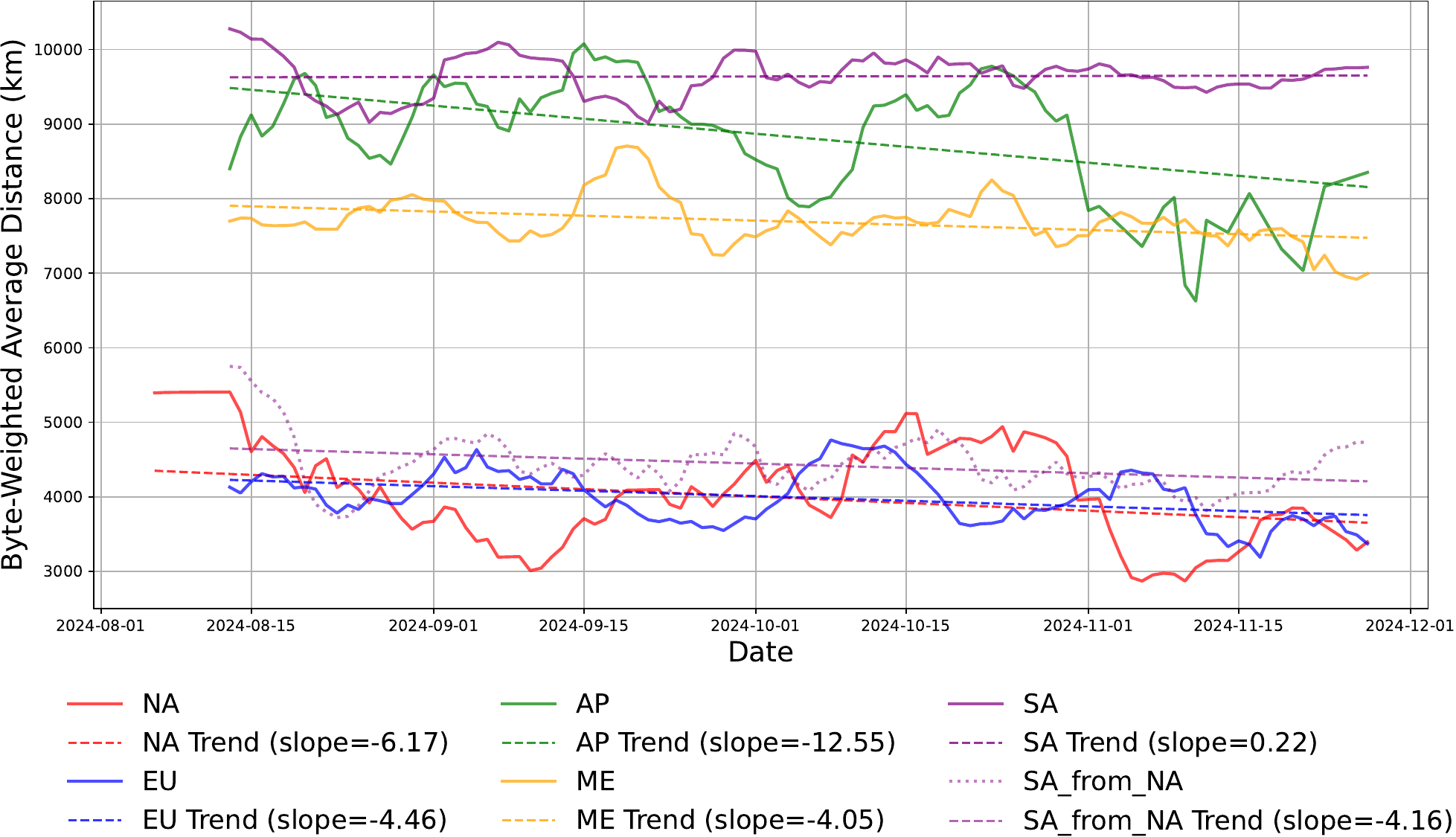}
    \caption{Daily byte-weighted average distance to peers for each vantage point over 107 days, with linear regression and 7-day smoothing.}
    \label{fig:distance}
\end{figure}

\begin{figure}[t]
    \centering
    \includegraphics[width=\linewidth]{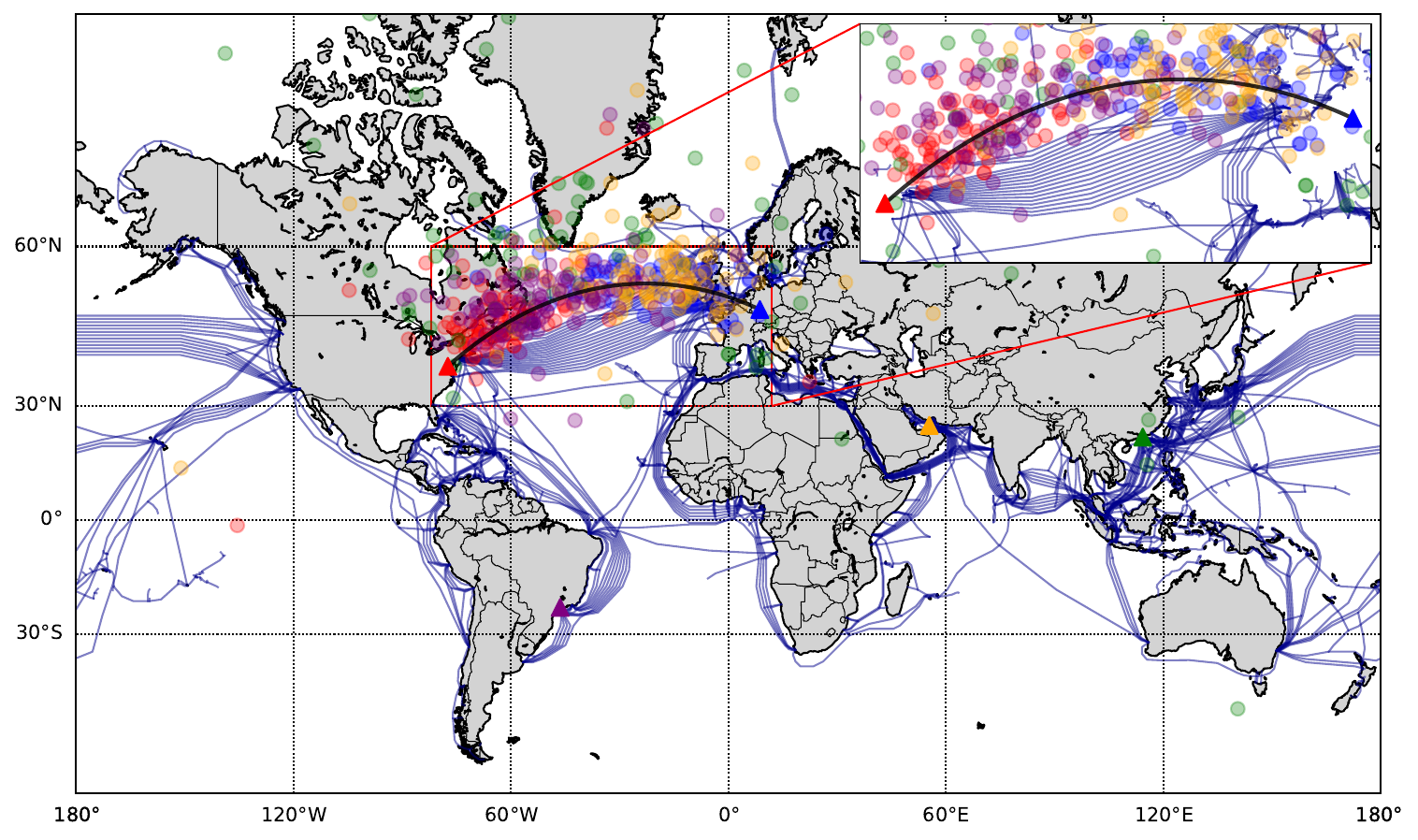}
    \caption{Data-weighted geographic focal points of five Ethereum nodes over 107 days. The black line is the geodesic path between N. Virginia and Frankfurt; blue lines show key undersea cables, highlighting convergence along critical routes.}
    \label{fig:focal}
\end{figure}

Four of the five vantage points showed a significant shift toward lower-latency peers. For example, the North American node’s average peer distance shrank by 6.17\,km/day ($p=0.002$), indicating intensified geographic clustering (\cref{fig:distance}). The South American node did not exhibit local centralization ($p=0.797$), but re-centering it as if it were in North America produced a strong clustering effect ($-4.16$\,km/day, $p<0.001$), reflecting sparse local peers and dependence peer connections in NA, afforded by undersea cables connecting the regions. Daily weighted focal points converged along NA--EU submarine routes (\cref{fig:focal}), underscoring reliance on limited physical links. Meanwhile, nodes in remote regions were pruned more often, reinforcing existing hubs and reducing regional connectivity.

These results support the hypothesis that \textbf{pruning based on latency unintentionally drives geographic centralization}, raising resilience and censorship concerns. Reliance on a handful of undersea cable routes concentrates risk: any disruption along these paths can disproportionately affect large swaths of the network. Balancing performance with broader peer diversity will require refined scoring mechanisms that look beyond latency alone, ultimately safeguarding Ethereum’s decentralization at the P2P layer.

\section{Conclusion and Data Release}

We introduced \name, a global dataset combining Ethereum node metrics, honeypot data, and full network traffic sessions from ten vantage points over three months, enabling holistic analyses of node performance, P2P topology, and security threats. Our findings show how client-based networking decisions can inadvertently drive geographic centralization, affecting resilience and censorship resistance. The complete \(\sim\)16\,TB compressed dataset is available upon request and other access options are available on OSF at \url{https://doi.org/10.17605/OSF.IO/C5UPF}.

\section{Acknowledgments}

The funding for this dataset was provided by the Ethereum Foundation under grants FY24-1546 and FY24-1547.

\bibliographystyle{IEEEtran}
\bibliography{references}

\end{document}